\documentclass[reprint,superscriptaddress,
showpacs,
 amsmath,amssymb,
 aps,
pra,
]{revtex4-1}

\usepackage{graphicx}
\usepackage{dcolumn}
\usepackage{bm}

\usepackage{epstopdf}
\usepackage{tikz}
\usepackage[normalem]{ulem}

\newcommand{\kT}{k_{\rm B}T}
\newcommand{\md}{\mathrm{d}}

\newcommand{\TLM}{{\mathcal L}(\Lambda)}
\newcommand{\TLMA}{{\mathcal L}(\langle\Lambda\rangle_{\Omega})}
\newcommand{\mcP}{\mathcal P}
\newcommand{\SubScr}{\Omega}

\begin{document}


\title{Stochastic control in microscopic nonequilibrium systems}

\author{Steven J. Large}
\email{slarge@sfu.ca}
\affiliation{Department of Physics, Simon Fraser University, Burnaby, British Columbia, V5A 1S6 Canada}

\author{Rapha\"{e}l Chetrite}
\affiliation{Laboratoire J.A. Dieudonn\'{e}, UMR CNRS 6621, Universit\'{e} de Nice Sophia-Antipolis, Nice 06108, France}
\affiliation{Pacific Institute of Mathematical Sciences, UMI 3069, Vancouver, British Columbia, Canada}

\author{David A. Sivak}%
\email{dsivak@sfu.ca}
\affiliation{Department of Physics, Simon Fraser University, Burnaby, British Columbia, V5A 1S6 Canada}

\date{\today}

\begin{abstract} 
Quantifying energy flows at nanometer scales promises to guide future research in a variety of disciplines, from microscopic control and manipulation, to autonomously operating molecular machines. 
A general understanding of the thermodynamic costs of nonequilibrium processes would illuminate the design principles for energetically efficient microscopic machines. 
Considerable effort has gone into finding and classifying the deterministic control protocols that drive a system rapidly between states at minimum energetic cost. 
But when the nonequilibrium driving is imposed by a molecular machine that is itself strongly fluctuating, driving protocols are stochastic.
Here we generalize a linear-response framework to incorporate such protocol variability and find a lower bound on the work that is realized at finite protocol duration, far from the quasistatic limit.
Our findings are confirmed in model systems.
This theory provides a thermodynamic rationale for rapid operation, independent of functional incentives.  
\end{abstract}


\maketitle

\section{\label{sec:introduction}Introduction}
In the past two decades, significant strides have been made in uncovering the physics of nonequilibrium processes~\cite{seifert_2012,jarzynski_2011}. 
The fluctuation theorems, for instance, place stringent constraints on the behavior of physical systems even far from equilibrium~\cite{jarzynski_1997,crooks_1999,gallavotti_1995,evans_1994,evans_1993,hatano_2001}.  Complementary to theoretical progress, the development of a multitude of experimental techniques to probe the microscopic physics of fluctuating systems has led to the direct verification of these strikingly general descriptions of the fluctuations and dissipation in physical systems~\cite{toyabe_2010,collin_2005,wang_2002,carberry_2004,garnier_2005}.  

While the fluctuation theorems characterize general properties of thermodynamic systems, they don't directly address questions of optimality.
For instance, there
is great interest in studying the efficiency of driven nonequilibrium systems, toward the goal of understanding the physical limits of biomolecular processes, perhaps pointing to design principles~\cite{Brown:2017ta}. 
A paradigmatic model system is the ${\rm F}_{\rm o}{\rm F}_1$ ATP synthase rotary motor, which uses rapid (presumably far-from-equilibrium) mechanical rotation of a crankshaft---itself driven by proton flow across a membrane---to drive synthesis of ATP molecules~\cite{yoshida_2001}.
We hypothesize that evolution has placed selective pressure on the development of energetically efficient machinery~\cite{Niven:2008ki}, which suggests that uncovering general features of efficient nonequilibrium driving may shed light on the fundamental principles underlying the design of microscopic machines.
Better understanding of such biomolecular machines promises practical benefits ranging from the \textit{de novo} construction of synthetic motors for next-generation nanomedicine~\cite{browne_2006} to a better understanding of diseases related to cellular transport, such as ALS and Alzheimer's~\cite{mandelkow_2002}.
		
To address these questions, we adopt a framework quantifying the nonequilibrium efficiency of time-dependent driving protocols connecting the initial and final system macrostates~\cite{sivak_2012}.  This formalism has been applied to a number of model thermodynamic systems~\cite{Sivak:2016:PhysRevE,zulkowski_2015,zulkowski_2012,rotskoff_2015}, and promises to inform the design of future single-molecule experiments on biophysical systems~\cite{gore_2003}.  

Efforts in this area have focused on deterministic protocols in experimental paradigms such as flipping or erasing a classical bit~\cite{Zulkowski:2014:PhysRevE}, or manipulating a biomacromolecule using optical traps or atomic force microscopy~\cite{Sivak:2016:PhysRevE}. 
A deterministic protocol lends itself naturally to single-molecule experiments, where the same time-dependent driving protocol can be reliably repeated. 
Yet in biomolecular contexts, the nonequilibrium driving may be imposed by molecular machines that are composed of protein components. At ambient temperature, these soft-matter system components (such as the crankshaft of ATP synthase) undergo strong conformational fluctuations, hence can only provide stochastic driving protocols to downstream systems (such as the ${\rm F}_1$ subunit that synthesizes ATP).
In order to probe the thermodynamics of stochastic driving in autonomous systems, we consider energetic costs that arise from a statistical distribution of control protocols.  

In this paper we generalize the linear-response formalism from \cite{sivak_2012} so that it quantifies energetic costs associated with statistical ensembles of control protocols.  
Our central result is that this variation in control protocols creates an additional energetic cost associated with slow operation, leading to work being minimized at finite protocol duration.  
Under the linear-response approximation, the lower bound on work~\eqref{lower_bound} and optimal duration~\eqref{optimal_time} take on simple forms.
For a single control parameter operating within these limits with uniform friction coefficient and control parameter velocity fluctuations, this implies an optimal mean driving velocity equal to the standard deviation of those stochastic velocity fluctuations~\eqref{optimal_velocity_uniform}. 
Our theoretical formulation identifies the existence of a minimal cost for stochastic control -- the only control modality available for living soft-matter systems.

\section{\label{sec:background}Theoretical background}
We consider a system in contact with a heat bath at temperature $T$, with equilibrium distribution
\begin{equation}
\pi(x|\boldsymbol{\lambda}) = e^{-\beta E(x,\boldsymbol{\lambda}) + \beta F(\boldsymbol{\lambda})}\label{boltzmann} \ ,
\end{equation}
over microstates $x$ with energy $E(x,\boldsymbol{\lambda})$ given experimentally controlled parameters $\boldsymbol{\lambda}$. Here $F(\boldsymbol{\lambda})$ is the equilibrium free energy and $\beta \equiv (\kT)^{-1}$ the inverse temperature.
A control \emph{protocol} $\Lambda: \boldsymbol{\lambda}_{\rm i}\rightarrow\boldsymbol{\lambda}_{\rm f}$ is a schedule of changing the control parameters $\boldsymbol{\lambda}(t)$ from an initial $\boldsymbol{\lambda}_{\rm i}$ at $t=0$ to a final $\boldsymbol{\lambda}_{\rm f}$ at time $\tau$.  
$W_{\rm ex} \equiv W - \Delta F$ is the excess work expended in performing protocol $\Lambda$, i.e. the work required above and beyond the equilibrium free energy change $\Delta F$.   

Within the linear-response regime, the excess power at time $t$ in a given control protocol $\Lambda$, averaged over system responses, takes on the integral expression~\cite{sivak_2012}
\begin{equation}
    \langle \mathcal{P}_{\rm ex}\rangle_{\Lambda_{t}} = \dot{\lambda}^i(t)\int_{-\infty}^{t}\langle\delta f_i(0)\delta f_j(t - t')\rangle_{\boldsymbol{\lambda}(t)}\dot{\lambda}^j(t')\md t' \label{excess_power_pre-integration} \ ,
\end{equation}
where $\dot{\lambda}^i = \md\lambda^i/\md t$ denotes differentiation with respect to time, angled brackets $\langle\cdots\rangle_{\Lambda_{t}}$ indicate an instantaneous average at time $t$ over system responses to protocol $\Lambda$, angled brackets $\langle\cdots\rangle_{\boldsymbol{\lambda}(t)}$ indicate an average over equilibrium fluctuations at fixed control parameters $\boldsymbol{\lambda}(t)$, and $f_i \equiv -\partial_{\lambda^i} E$ is the generalized force conjugate to the $i$th control parameter.  Throughout we employ the Einstein summation notation, implicitly summing over any repeated indices.

If the control protocol $\Lambda$ is sufficiently smooth, such that
\begin{equation}
    \dot{\lambda}^j(t) \gg (t' - t) \ddot{\lambda}^j(t) \label{protocol_constraint} \ ,
\end{equation}
for time separations $t' - t$ over which the conjugate force autocorrelation $\langle\delta f_i(0)\delta f_j(t-t')\rangle_{\boldsymbol{\lambda(t)}}$ is significantly greater than zero, then the $j$th control parameter velocity in~\eqref{excess_power_pre-integration} can be approximated by its current value, $\dot{\lambda}^j(t')\approx \dot{\lambda}^j(t)$, and the excess power becomes
\begin{equation}
    \langle \mathcal{P}_{\rm ex}\rangle_{\Lambda_t} = \dot{\lambda}^i\zeta_{ij}(\boldsymbol{\lambda})\dot{\lambda}^j \label{excess_power} \ .
\end{equation}
In what follows, for notational convenience we suppress the explicit time dependence of $\dot{\lambda}^i$ (see SI section~V for more details). 

$\zeta_{ij}(\boldsymbol{\lambda})$ is a generalized friction tensor on the space of control parameters,
\begin{equation}
    \zeta_{ij}(\boldsymbol{\lambda}) \equiv \beta\int_0^{\infty}\langle\delta f_i(0)\delta f_j(t)\rangle_{\boldsymbol{\lambda}}\md t = \beta \tau_{ij}^{\rm R}\langle\delta f_i\delta f_j\rangle_{\boldsymbol{\lambda}} \ ,
\end{equation}
where $\tau_{ij}^{\rm R}$ is the integral relaxation time~\cite{Garanin:1996:PhysRevE}, and
$\langle\delta f_i\delta f_j\rangle_{\boldsymbol{\lambda}}$ is the equilibrium force variance~\cite{sivak_2012}.
Under linear response, the excess work is
\begin{equation}
\langle W_{\rm ex}\rangle_{\Lambda} = \int_0^{\tau}\langle \mcP_{\rm ex}\rangle_{\Lambda_t} \md t \approx \int_{0}^{\tau}\dot{\lambda}^i\zeta_{ij}(\boldsymbol{\lambda})\dot{\lambda}^j \md t\label{excess_work_deterministic} \ ,
\end{equation}
for duration $\tau$ of the control protocol $\Lambda$.

The generalized friction tensor $\zeta_{ij}(\boldsymbol{\lambda})$ also provides a measure of thermodynamic length~\cite{crooks_2007}
\begin{equation}
\TLM \equiv \int_{0}^{\tau}\sqrt{\dot{\lambda}^i\zeta_{ij}(\boldsymbol{\lambda})\dot{\lambda}^j} \, \md t\label{thermodynamic_length} \ ,
\end{equation}
along a protocol $\Lambda$. 
For a given path in control parameter space connecting $\boldsymbol{\lambda}_{\rm i}$ to $\boldsymbol{\lambda}_{\rm f}$, the thermodynamic length is independent of the protocol duration $\tau$,
and through a Cauchy-Schwarz inequality provides a lower bound on the excess work, $\langle W_{\rm ex}\rangle_{\Lambda} \geq \TLM^2/\tau$. This bound is saturated for a given duration $\tau$ when the protocol follows the geodesic curve connecting $\boldsymbol{\lambda}_{\rm i}$ to $\boldsymbol{\lambda}_{\rm f}$~\cite{crooks_2007,ruppeiner_1979,nulton_1985,salamon_1983,nulton_1984,schlogl_1985,Schmiedl_2007,andreson_1984}.

\section{\label{sec:protocol_ensembles}Protocol ensembles and stochastic control}
Here, instead of a single protocol $\Lambda:\boldsymbol{\lambda}_{\rm i}\rightarrow\boldsymbol{\lambda}_{\rm f}$, we consider an ensemble $\Omega$ of protocols, where each protocol $\Lambda$ satisfies~\eqref{protocol_constraint} and occurs with probability $P[\Lambda|\Omega]$. The excess power $\langle \mathcal{P}_{\rm ex}\rangle_{\Lambda_t}$ at time $t$ 
during protocol $\Lambda\in\Omega$, averaged over system fluctuations,  
is now a random variable~\eqref{excess_power} since $\dot{\lambda}^i$, $\zeta_{ij}(\boldsymbol{\lambda})$, and $\dot{\lambda}^j$ are all functions of the random protocol $\Lambda$. 
The excess power, averaged over system \emph{and} protocol fluctuations, is
\begin{equation}
\langle \mcP_{\rm ex}\rangle_{\Omega_t} \equiv \int \langle \mcP_{\rm ex}\rangle_{\Lambda_t}P[\Lambda|\Omega] \, \mathcal{D}[\Lambda] \ ,
\label{general_excess_power}
\end{equation}
where the integral is taken over all protocols and hence all instantaneous values of $\dot{\boldsymbol{\lambda}}(t)$.
$\langle \cdots\rangle_{\Omega_t}$ indicate an average over the instantaneous distribution of control parameter positions or velocities at time $t$ due to the protocol ensemble $\Omega$.

When the ensemble is tightly localized around the average protocol, such that the friction varies little  over control parameter values with significant support, the excess power (averaged over protocol and system fluctuations) is well approximated by expanding $\langle \mcP_{\rm ex}\rangle_{\Lambda_t}$~\eqref{excess_power} about the mean values of its arguments $\dot{\lambda}^i$, $\dot{\lambda}^j$, and $\zeta_{ij}(\boldsymbol{\lambda})$~\cite{casella_2002}:  
\begin{equation}
\langle \mcP_{\rm ex}\rangle_{\Omega_t} \approx \langle\dot{\lambda}^i\rangle_{\Omega_t}\zeta_{ij}(\langle\boldsymbol{\lambda}\rangle_{\Omega_t})\langle\dot{\lambda}
^j\rangle_{\Omega_t} + \zeta_{ij}(\langle\boldsymbol{\lambda}\rangle_{\Omega_t})\langle\delta\dot{\lambda}^i\delta\dot{\lambda}^j\rangle_{\Omega_t}\label{excess_power_stochastic} \ .
\end{equation}
SI section~I gives a full derivation using a weak-noise perturbation expansion.

Time integration of \eqref{excess_power_stochastic} gives the average excess work required to perform a random protocol sampled from $\Omega$,
\begin{align}
&\langle W_{\rm ex}\rangle_{\Omega} = \label{excess_work_stochastic} \\
&\int_{0}^{\tau}\left[\langle\dot{\lambda}^i\rangle_{\Omega_t}\zeta_{ij}(\langle\boldsymbol{\lambda}\rangle_{\Omega_t})\langle\dot{\lambda}^j\rangle_{\Omega_t} + \zeta_{ij}	(\langle\boldsymbol{\lambda}\rangle_{\Omega_t})\langle\delta\dot{\lambda}^i\delta\dot{\lambda}^j\rangle_{\Omega_t} \right] \md t \ , \nonumber
\end{align}
where $\langle\cdots\rangle_{\Omega}$ indicates an average over all protocols $\Lambda\in\Omega$, weighted by $P[\Lambda|\Omega]$.
The first RHS term resembles~\eqref{excess_work_deterministic}, quantifying the cost associated with fast operation, while the second term quantifies the energetic cost resulting from variability in the protocol velocities. 
Both terms are integrated along the (deterministic) average protocol specified by the average velocity $\langle\dot{\boldsymbol{\lambda}}\rangle_{\Omega_t}$.
Thus, in the weak protocol-noise limit the effect of variable control only depends on the friction along this average path and the variation in velocities as a function of time. 

\subsection{\label{sec:lower_bound}Lower bound on excess work}
The Cauchy-Schwarz inequality gives a lower bound for the first RHS term in \eqref{excess_work_stochastic} involving the thermodynamic length $\TLMA$ between the initial and final states of the average protocol $\langle\Lambda\rangle_{\Omega}$~\cite{crooks_2007}, leading to a lower bound on the excess work achieved at a finite protocol duration $\tau$,
\begin{equation}
\langle W_{\rm ex}\rangle_{\Omega} \geq \frac{\TLMA^2}{\tau} + \left\langle \zeta_{ij}(\langle\boldsymbol{\lambda}\rangle_{\Omega_t})\langle\delta\dot{\lambda}^i\delta\dot{\lambda}^j\rangle_{\Omega_t}\right\rangle_{\SubScr} \tau \ ,
\label{excess_work_final}
\end{equation}
where we write the average of an instantaneous quantity over the protocol ensemble as $\langle \cdots \rangle_{\Omega} \equiv \tau^{-1} \int_0^{\tau} \cdots \, \md t$. 

This lower bound represents a tradeoff between the first RHS term quantifying the energetic costs associated with pushing a system out of equilibrium (scaling as $\tau^{-1}$ with protocol duration) and the second term quantifying the average contribution of protocol fluctuations to excess work, which increases with $\tau$ if $\zeta_{ij}(\boldsymbol{\lambda})$ is positive definite (assumed in what follows).

If the control parameter velocity variance is independent of the average velocity, this lower bound is minimized at a finite protocol duration 
\begin{align}
\label{optimal_time}
\tau^{\rm opt} &= \frac{\TLMA}{\left\langle\zeta_{ij}(\langle\boldsymbol{\lambda}\rangle_{\Omega_t})\langle\delta\dot{\lambda}^i\delta\dot{\lambda}^j\rangle_{\Omega_t}\right\rangle_{\SubScr}^{1/2}}\ ,
\end{align}
revealing a fundamental lower bound on the excess work,
\begin{align}
\label{lower_bound} 
\langle W_{\rm ex}\rangle_{\Omega} &\geq 2\left\langle \zeta_{ij}(\langle\boldsymbol{\lambda}\rangle_{\Omega_t})\langle\delta\dot{\lambda}^i\delta\dot{\lambda}^j\rangle_{\Omega_t}\right\rangle_{\SubScr}^{1/2}\TLMA \ .
\end{align}
This lower bound is saturated when the average protocol $\langle\Lambda\rangle_{\Omega}$ follows the geodesic from $\boldsymbol{\lambda}_{\rm i}$ to $\boldsymbol{\lambda}_{\rm f}$ (similar to the deterministic case).
The existence of a lower bound on work realized at finite protocol duration constitutes our main result. In the linear-response regime, the lower bound and optimal protocol duration take on the simple forms in~\eqref{lower_bound} and~\eqref{optimal_time}, respectively.

For a single control parameter, \eqref{lower_bound} and \eqref{optimal_time} can be recast solely in terms of intensive quantities as a lower bound on the average excess force $\langle f_{\rm ex}\rangle_{\Omega} \equiv \langle W_{\rm ex}\rangle_{\Omega}/\Delta\lambda$
produced by an optimal mean control parameter velocity $\langle\dot{\lambda}\rangle^{\rm opt}_{\Omega} \equiv \Delta\lambda/\tau^{\rm opt}$.
When the friction and control-parameter velocity variance are both uniform, the excess force bound 
and optimal velocity 
simplify to
\begin{subequations}
\begin{align}
\langle\dot{\lambda}\rangle^{\rm opt}_{\Omega} &= \sqrt{\langle\delta\dot{\lambda}^2\rangle_{\Omega}} \label{optimal_velocity_uniform} \\
\langle f_{\rm ex}\rangle_{\Omega} &\geq 2\zeta\langle\dot{\lambda}\rangle^{\rm opt}_{\Omega} \label{lower_bound_force_uniform} \ .
\end{align}
\end{subequations}
The optimal mean velocity is the root-mean-squared control-parameter velocity fluctuations, producing a mean excess force equal to twice the Stokes drag on the control parameter when moving at the optimal mean velocity through the `viscous' control parameter space subject to generalized friction coefficient $\zeta$.

In the specific case in which across the entire protocol the integral relaxation time is constant and equals
$\tau^{\rm R} = (\beta \langle\zeta_{ij}(\langle\boldsymbol{\lambda}\rangle_{\Omega_t})\langle\delta\dot{\lambda}^i\delta\dot{\lambda}^j\rangle_{\Omega_t}\rangle_{\SubScr})^{-1}$, our lower bound~\eqref{lower_bound} reduces to Machta's bound on entropy production of a stochastically driven process~\cite{machta_2015}.
This equality is achieved in the one-dimensional drift-diffusion process considered by Machta when protocol fluctuations come from the interaction of the control parameter with a thermal reservoir at the same temperature as the reservoir producing system fluctuations.
Thus our derived lower bound~\eqref{lower_bound} generalizes Machta's bound to systems with variable integral relaxation times and arbitrary fluctuations of the control parameter. 
SI section~II gives more details.

\section{\label{model}Model ensembles}
\begin{figure}
\centering\includegraphics[width=\columnwidth]{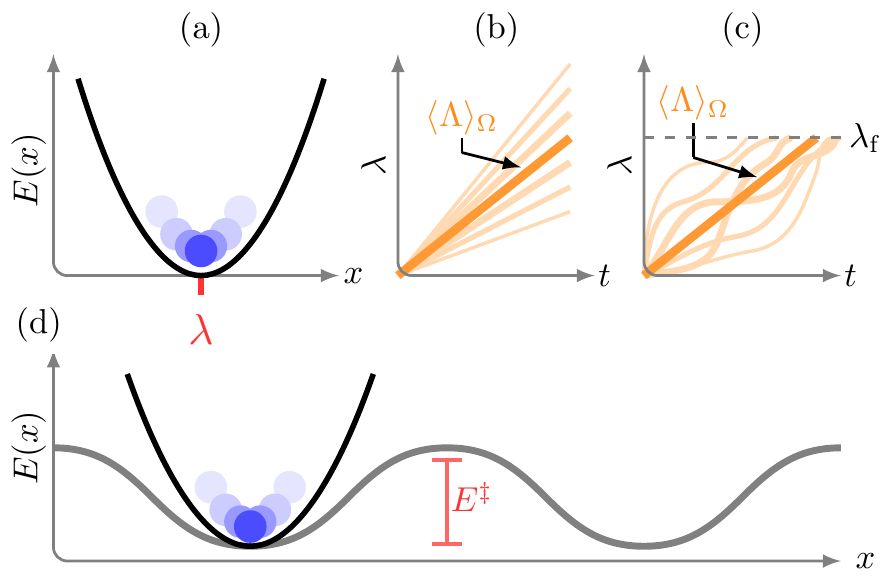}
\caption{{\bf Schematic of model system and protocol ensembles.}  (a) Brownian particle diffusing in a harmonic potential, with harmonic trap minimum the control parameter.
(b,c) Protocol samples from the zero-barrier periodic potential ensemble and the stochastically driven ensemble. 
Bold lines denote average protocols. (d) Energy landscape for periodic potential with barrier height $E^{\ddagger}$ between adjacent energetic minima.
\label{model_figure}}
\end{figure}

We illustrate our theoretical approximation~\eqref{excess_work_final} using two model protocol ensembles.  In each case, the system is a Brownian particle with unit mass evolving according to an overdamped Langevin equation on a one-dimensional potential. Driving forces are produced by a harmonic potential $U(x,\lambda) = \tfrac{1}{2}k[x - \lambda(t)]^2$, with trap strength $k$ and control parameter $\lambda(t)$ the time-dependent potential minimum (Fig.~\ref{model_figure}). To saturate the excess work bounds in (\ref{excess_work_final},\ref{lower_bound}), we restrict our attention to protocol ensembles where the average protocol $\langle\Lambda\rangle_{\Omega}$ is the minimum-work protocol~\cite{sivak_2012}. SI section~III provides simulation details for each ensemble. 

For one control parameter,
the theoretical minimum excess work for an ensemble $\Omega$ of driving protocols operating within the linear-response regime~\eqref{excess_work_final} with a constant control parameter velocity variance simplifies to
\begin{equation}
\langle W_{\rm ex}\rangle_{\Omega} \geq \frac{\TLMA^2}{\tau} + \langle\zeta(\lambda)\rangle_{\Omega} \langle \delta\dot{\lambda}^2\rangle_{\Omega} \, \tau \ .
\label{work_harmonic}
\end{equation}

\subsection{\label{sec:constant_velocity}Periodic-potential ensemble}

Here
the harmonic trap is driven over an underlying periodic potential $U_{\rm period}(x) = -\tfrac{1}{2}E^{\ddagger}\cos \pi x$ with energy barrier $E^{\ddagger}$ between adjacent wells (Fig.~\ref{model_figure}d).
Similar potentials have been used to investigate systems with a sequence of metastable states, which are popular models of the basic physics of molecular machines~\cite{reimann_2002}.
The generalized friction $\zeta(\lambda)$ can be expressed as~\cite{zulkowski_2015,berezhkovskii_2011}
\begin{equation}
\zeta(\lambda) = \frac{1}{\beta D}\int_{-\infty}^{\infty}\frac{\left[ \Pi_{\rm eq}(x|\lambda) \right]^2}{\pi(x|\lambda)}\md x \ ,
\label{zulkowski_friction}
\end{equation}
for equilibrium cumulative distribution function $\Pi_{\rm eq}(x|\lambda) \equiv \int_{-\infty}^{x} \pi(x'|\lambda)\md x'$ and system diffusion coefficient $D$.

We examine a protocol ensemble where each protocol $\Lambda$ completes the minimum-work path with an average velocity $\langle\dot{\lambda}\rangle_{\Lambda}$ randomly sampled from a Gaussian distribution with mean $\langle\dot{\lambda}\rangle_{\Omega}$ and variance $\langle\delta\dot{\lambda}^2\rangle_{\Omega}$.
Each protocol has instantaneous velocity $\dot{\lambda}\propto\left[\zeta(\lambda)\right]^{-1/2}$ with the proportionality fixed by the prescribed average velocity $\langle\dot{\lambda}\rangle_{\Lambda}$.
The ensemble-mean control parameter velocity $\langle\dot{\lambda}\rangle_{\Omega} = \langle\Delta\lambda\rangle_{\Omega}/\tau$ is chosen so that the average protocol $\langle\Lambda\rangle_{\Omega}$ completes the control parameter change $\langle\Delta\lambda\rangle_{\Omega}$ in a prescribed time $\tau$.  
The system is initialized in the periodic 
steady state for a harmonic trap traversing the periodic minimum-work protocol at the particular chosen average velocity.

In the zero-barrier limit, the friction is constant, so the minimum-work protocol proceeds with a constant velocity,  
thus
producing an exact mean excess work~\eqref{excess_work_final}
\begin{equation}
\langle W_{\rm ex}\rangle_{\Omega} = \zeta\langle\dot{\lambda}^2\rangle_{\Omega} \, \tau
= \frac{\TLMA^2}{\tau} + \zeta\langle\delta\dot{\lambda}^2 \rangle_{\Omega} \, \tau  \ , 
\label{ConstVel_exact}
\end{equation}
because $\langle \dot{\lambda}^2\rangle_{\Omega} = \langle\dot{\lambda}\rangle_{\Omega}^2 + \langle\delta\dot{\lambda}^2\rangle_{\Omega}$ and for this system, where the control parameter velocity variance $\langle\delta\dot{\lambda}^2\rangle_{\Omega}$ is constant across all $\lambda$, \eqref{thermodynamic_length} simplifies to $\TLMA = \sqrt{\zeta}\langle\dot{\lambda}\rangle_{\Omega}\, \tau$.
(SI section~IV provides a full derivation.)

\begin{figure}[t!]
\includegraphics[width=\columnwidth]{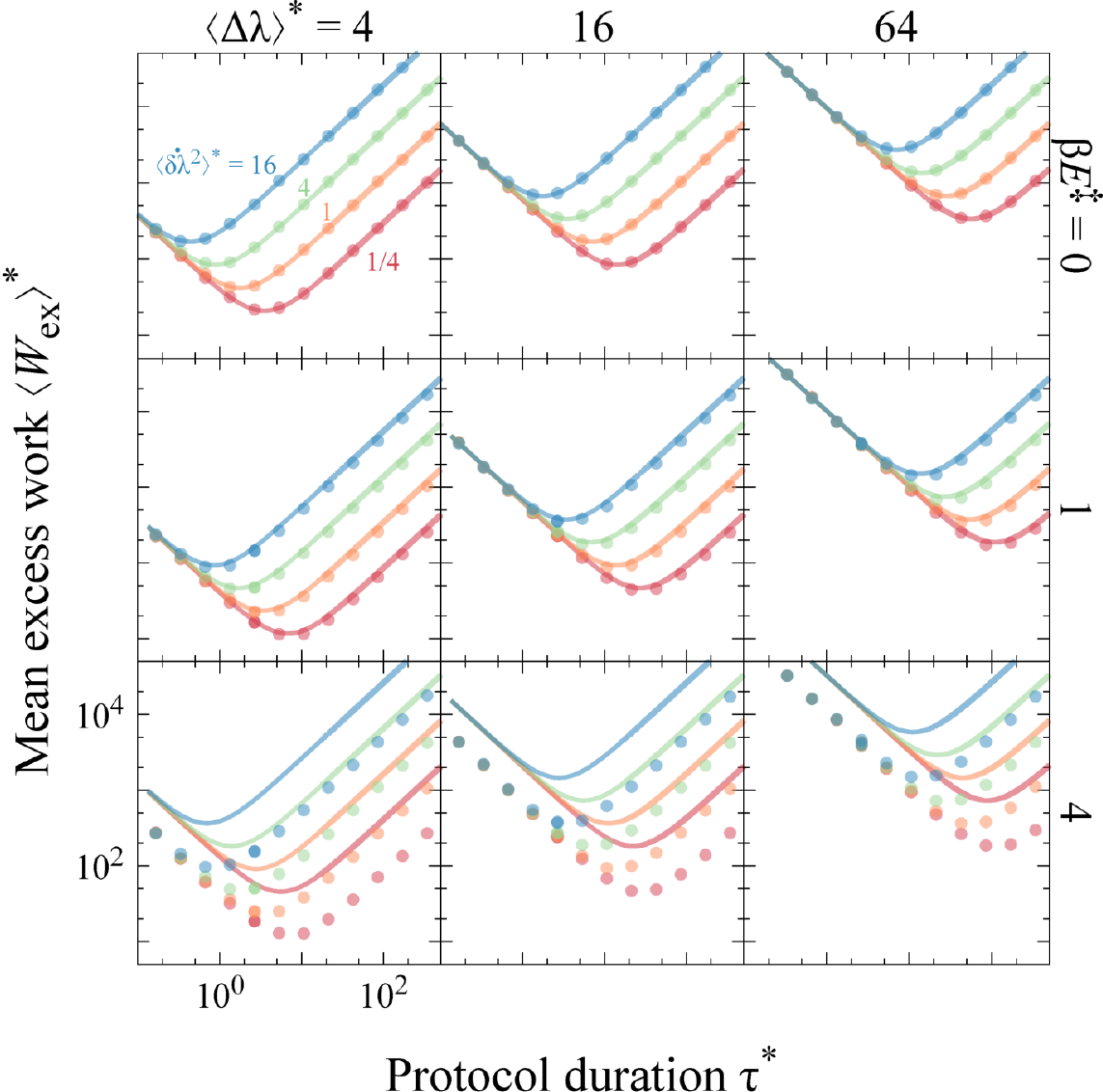}
\caption{{\bf Excess work for periodic-potential 
ensemble is minimized at finite protocol duration.}
Excess work $\langle W_{\rm ex}\rangle_{\Omega}^* \equiv \beta\langle W_{\rm ex}\rangle_{\Omega}$ (in units of thermal energy), 
as a function of protocol duration $\tau^* \equiv \tau/(2\beta Dk)^{-1}$ (scaled by the time taken to diffuse the standard deviation $\sqrt{\langle \delta x^2 \rangle_{\lambda}} \equiv (\beta k)^{-1/2}$ in equilibrium position).
The protocol distance $\lambda^* \equiv \lambda/(\beta k)^{-1/2}$ is scaled by $\sqrt{\langle \delta x^2 \rangle_{\lambda}}$. 
Nondimensionalized control parameter velocity variance $\langle\delta\dot{\lambda}^2\rangle^*\equiv \langle\delta\dot{\lambda}^2\rangle/(4\beta D^2k)$ ranges from high (blue) to low (red).  Each row shows a different periodic barrier height $\beta E^{\ddagger}$ from $0$ (top) to $4$ (bottom).  The underlying potential has spatial periodicity $L = 4\sqrt{\langle\delta x^2\rangle_{\lambda}}$. 
\label{ConstantVelocityEnsemble}}
\end{figure}

Figure~\ref{ConstantVelocityEnsemble} shows a comparison of numerical calculations to theoretical predictions for several average protocol distances and periodic barrier heights.  For no underlying barrier ($\beta E^{\ddagger}$ = 0), the numerical data agrees exactly with the analytical solution.  
$\beta E^{\ddag} = 1$ shows good agreement, but
with increasing barrier height, the linear-response approximation in~\eqref{protocol_constraint} begins to break down for rapid protocols, 
and the numerical results depart from the theoretical predictions.  However, for all barriers explored, even those for which~\eqref{protocol_constraint} does not hold, the excess work is minimized at finite protocol duration.

\subsection{\label{sec:stochastic_control}Stochastically driven protocols}

Here
the protocol itself evolves 
according to a dynamic stochastic process,
traveling between given initial and final control parameter values $\lambda_{\rm i}$ and $\lambda_{\rm f}$ in a variable duration $\tau'$.  
The system is initialized in the nonequilibrium steady-state (NESS) for the harmonic trap moving with the (constant) average velocity of the protocol ensemble.

The control parameter dynamics obey an underdamped Langevin equation~(S34)
with potential energy $U_{\lambda}(\lambda,\lambda_0(t)) = \tfrac{1}{2} k_{\lambda} [\lambda - \lambda_0(t)]^2$ that is harmonic with spring constant $k_{\lambda}$ confining the control parameter and time-dependent minimum $\lambda_0(t)$. 
$\lambda_0(t)$ moves with constant velocity, and throughout the 
protocol the distribution of control parameter positions and velocities is stationary in the frame which is comoving with $\lambda_0(t)$.
As a result, the average control parameter velocity is constant, and the average protocol $\langle\Lambda\rangle_{\Omega}$ is the minimum-work protocol~\cite{sivak_2012}. 
The steady-state variance of the control parameter velocity is fixed in the comoving frame by the equipartition theorem, $\langle\delta\dot{\lambda}^2\rangle_{\Omega_t} = (\beta_{\lambda}m_{\lambda})^{-1}$, where $m_{\lambda}$ is the mass of the control parameter~\cite{vankampen}.  
If control parameter velocity fluctuations persist over time scales longer than the system relaxation time, then~\eqref{protocol_constraint} holds for all stochastic protocols in the ensemble.
Effectively, this represents a locally deterministic limit, where over relaxation time scales of the system, the control parameter is largely unaffected by stochastic fluctuations, but still exhibits large fluctuations over longer time scales (see SI section~V for details).

\begin{figure}[h!]
\includegraphics[width=\columnwidth]{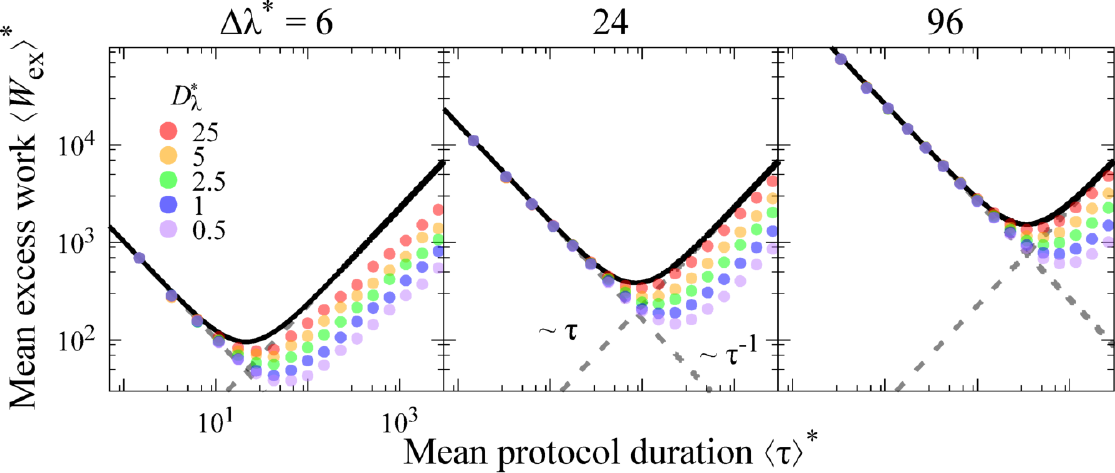}
\caption{{\bf Excess work for stochastic protocol ensemble matches theoretical approximation in the locally deterministic limit.}
Mean excess work $\langle W_{\rm ex}\rangle^*$ as function of mean protocol duration $\langle \tau \rangle^*$.  For underdamped control parameter dynamics and large protocol distances (where control parameter dynamics are locally deterministic), numerical simulations (circles) agree with the theoretical approximation~\eqref{excess_work_final} (solid black curve), composed of terms proportional and inversely proportional to protocol duration (dashed black curves). 
Control parameter diffusion coefficient $D^*_{\lambda}\equiv D_{\lambda}/D$ (nondimensionalized by the system diffusion coefficient) interpolates between overdamped (purple) and underdamped (red) control parameter dynamics. 
}
\label{stochastic_control}
\end{figure}

Figure~\ref{stochastic_control} shows the average excess work as a function of the average protocol duration $\langle \tau \rangle_{\Omega}$, for several protocol distances $\Delta\lambda$ and control parameter diffusion coefficients $D_{\lambda}$.
Numerical simulations agree well with the theoretical predictions at short protocol durations where the excess work is dominated by the contribution from the average protocol~\eqref{work_harmonic}.
At long protocol durations, for intermediate to large protocol distances and high $D_{\lambda}$, the locally deterministic approximation~\eqref{protocol_constraint} 
is satisfied and the theoretical predictions agree well with the numerical results.
In all cases, the excess work is an increasing function of protocol duration in the long-duration limit.  
Thus, regardless of the theoretical approximation's accuracy, a finite-time lower bound on the excess work is widely observed, contrary to the case of deterministic protocols.

\section{\label{sec:discussion}Discussion}
In this letter, we present a formalism that generalizes previous theory to now quantify the nonequilibrium costs of driving a system with an ensemble of protocols.  
We assume only that the linear-response approximation applies for each protocol in the ensemble and that variation about the average protocol is sufficiently small.  
In these limits, protocol variation produces an additional energetic cost that increases with protocol duration.  

This theoretical framework gives rise to a lower bound on the excess work~\eqref{lower_bound} that generalizes a previous result~\cite{machta_2015} to arbitrary low-noise protocol ensembles and situations where the relaxation time varies across control parameter space.
Our expression for excess work makes transparent that the lower bound occurs for a finite protocol duration~\eqref{optimal_time} and hence finite average protocol velocity~\eqref{optimal_velocity_uniform}.
This implies an energetically optimal, finite time scale for the process, suggesting the novel possibility that biomolecular processes have energetically preferential time scales over which to operate, stemming from the statistical properties of their driving processes.

The resulting total work is completely specified by the average protocol and the variance of control parameter velocities~\eqref{excess_work_final}, so it may be identical for vastly different control strategies, each with potential advantages for particular tasks.  
This suggests that an autonomous system could simultaneously reduce the energetic cost of completing a particular thermodynamic process and improve an orthogonal quality metric through the clever choice of the statistical properties of the protocol ensemble 
(SI section~VI).

We have numerically investigated the consequences of these predictions in two model ensembles. Both the periodic-potential ensemble 
and the stochastic protocol ensemble 
show a finite-duration minimum for the excess work across all examined parameter space.
Complementary recent experiments~\cite{tafoya_2018} have shown that, even far from equilibrium, the linear-response formalism can be effective for reducing excess work in control protocols that unfold and refold 
a DNA hairpin.
Thus,
the qualitative trends predicted from our theoretical and numerical investigation may still prove insightful for the operational principles of biomolecular machines, even if such machines' natural operation quantitatively violates linear-response theory.

This theory is agnostic about the origin of such stochastic control parameter fluctuations, assigning work to any energy flow during control parameter changes.
Intriguing recent work~\cite{horowitz_2016} sheds light on the manner in which nonequilibrium reservoirs can perform work on thermodynamic systems and points toward more biophysically motivated models in which this theory could be applied. 
Recent research on strongly coupled systems~\cite{barato_2017} suggests connections with the framework developed here, so an open question for future work is the relation of our theory to a broader picture of multiple interacting stochastic systems~\cite{verley_2014}.

\section{Acknowledgements}

The authors thank Alzbeta Medvedova, Alexandra Kasper, Aidan Brown, Emma Lathouwers, John Bechhoefer (SFU Physics), Miranda Louwerse (SFU Chemistry), and Tomohiro Shitara (Univerity of Tokyo Physics) for insightful comments on the article.  This work is supported by Natural Sciences and Engineering Research Council of Canada (NSERC) CGS Masters and Doctoral fellowships (S.J.L.), an NSERC Discovery Grant (D.A.S.), a Tier II Canada Research Chair (D.A.S.), and WestGrid (www.westgrid.ca) and Compute Canada Calcul Canada (www.computecanada.ca).

%

\appendix

\section{\label{appendix:expansion_of_excess_power}Expansion of the excess power}

Within the linear-response regime, the instantaneous average excess power at time $t$ during protocol $\Lambda$ is $\langle \mcP_{\rm ex}\rangle_{\Lambda_t}\approx \dot{\lambda}^i\zeta_{ij}\dot{\lambda}^j$, where for notational convenience we suppress the dependence of $\zeta_{ij}$ on the control parameter.  For an ensemble $\Omega$ of protocols, where protocol $\Lambda$ occurs with probability $P[\Lambda|\Omega]$, the excess power at time $t$ (averaged over system responses to a given protocol) is 
itself stochastic.
We Taylor expand the linear-response approximation of the average excess power about its mean~\cite{casella_2002}:
\begin{align}
\langle \mcP_{\rm ex}\rangle_{\Lambda_t} &= \sum_{n,m,l}\frac{1}{n!m!l!} \times \label{appendix_expansion_general} \\
\partial_{\dot{\lambda}^i}^n\partial_{\zeta_{ij}}^m\partial_{\dot{\lambda}^j}^l \langle &\mcP_{\rm ex}\rangle_{\Lambda_t}\Big|_{\langle\dot{\lambda}^i\rangle_{\Omega_t},\langle\zeta_{ij}\rangle_{\Omega_t},\langle\dot{\lambda}^j\rangle_{\Omega_t}} (\delta\dot{\lambda}^i)^n(\delta\zeta_{ij})^m(\delta\dot{\lambda}^j)^l \ . \nonumber
\end{align}
This expansion requires that $\langle \mcP_{\rm ex}\rangle_{\Lambda_t}$ is a smooth function of $\dot{\lambda}^i$, $\dot{\lambda}^j$, and $\zeta_{ij}$, which clearly holds for $\langle \mcP_{\rm ex}\rangle_{\Lambda_t} = \dot{\lambda}^i\zeta_{ij}\dot{\lambda}^j$.

Keeping all nonzero terms, the excess power is
\begin{align}
\langle \mcP_{\rm ex}\rangle_{\Lambda_t} &= \langle\dot{\lambda}^i\rangle_{\Omega_t}\langle\zeta_{ij}\rangle_{\Omega_t}\langle\dot{\lambda}^j\rangle_{\Omega_t} + \langle\dot{\lambda}^i\rangle_{\Omega_t}\langle\dot{\lambda}^j\rangle_{\Omega_t}\delta\zeta_{ij} \nonumber \\ 
&+ \langle\dot{\lambda}^i\rangle_{\Omega_t} \langle\zeta_{ij}\rangle_{\Omega_t}\delta\dot{\lambda}^j + \langle\dot{\lambda}^j\rangle_{\Omega_t} \langle\zeta_{ij}\rangle_{\Omega_t} \delta\dot{\lambda}^j \nonumber \\
&+ \langle\dot{\lambda}^i\rangle_{\Omega_t} \delta\zeta_{ij}\delta\lambda^j + \langle\dot{\lambda}^j\rangle_{\Omega_t} \delta\zeta_{ij}\delta\dot{\lambda}^i \nonumber \\
&+ \langle\zeta_{ij}\rangle_{\Omega_t}\delta\dot{\lambda}^i\delta\dot{\lambda}^j + \delta\dot{\lambda}^i\delta\zeta_{ij}\delta\dot{\lambda}^j \label{appendix_expansion_explicit} \ ,
\end{align}
because terms of fourth and higher order are trivially zero by the form of $\langle \mcP_{\rm ex}\rangle_{\Lambda_t}$. 
Averaging $\langle \mcP_{\rm ex}\rangle_{\Lambda_t}$ over the protocol ensemble $\Omega$,
\begin{equation}
\langle \mcP_{\rm ex}\rangle_{\Omega_t} \equiv \int \langle \mcP_{\rm ex}\rangle_{\Lambda_t}P[\Lambda|\Omega]\mathcal{D}[\Lambda]\label{appendix_excess_power_functional_integral} \ ,
\end{equation}
all terms vanish trivially which are linear in (protocol) fluctuations from the mean, leaving: 
\begin{align}
\langle \mcP_{\rm ex}\rangle_{\Omega_t} &= \langle\dot{\lambda}^i\rangle_{\Omega_t}\langle\zeta_{ij}\rangle_{\Omega_t}\langle\dot{\lambda}^j\rangle_{\Omega_t} + \langle\dot{\lambda}^i\rangle_{\Omega_t}\langle\delta\zeta_{ij}\delta\dot{\lambda}^j\rangle_{\Omega_t} \nonumber \\ 
&+ \langle\dot{\lambda}^j\rangle_{\Omega_t}\langle\delta\zeta_{ij}\delta\dot{\lambda}^j\rangle_{\Omega_t}
+ \langle\zeta_{ij}\rangle_{\Omega_t}\langle\delta\dot{\lambda}^i\delta\dot{\lambda}^j\rangle_{\Omega_t} \nonumber \\ &+ \langle\delta\dot{\lambda}^i\delta\zeta_{ij}\delta\dot{\lambda}^j\rangle_{\Omega_t} \label{appendix_expansion_small} \ .
\end{align}

We assume the friction $\zeta_{ij}(\boldsymbol{\lambda})$ is a smooth function of the control parameter, which holds if all conjugate forces $f_i$ are even under momentum-reversal, except at a macroscopic phase transition.  
In the limit of weak noise~\cite{gardiner}, where the ensemble of protocols is tightly localized about its average, we expand the excess power perturbatively in noise strength.  Specifically, we assume that the $i,j$-th component of the friction tensor evolves in accordance with the general linear stochastic differential equation (SDE):
\begin{equation}
\dot{\zeta_{ij}}(\boldsymbol{\lambda},t) = a_{ij}(\boldsymbol{\zeta},t) + \epsilon b_{ij}^{\ell}(\boldsymbol{\zeta},t)\xi_{\ell}(t)
\label{appendix_general_linear_sde} \ ,
\end{equation}
where $\xi_{\ell}(t)$ is the $\ell$th element of a zero-mean (vector) white noise process affecting control parameter $\lambda^{\ell}$, with $\langle\xi_{\ell}(t)\xi_m(t')\rangle = \delta_{\ell,m}\delta(t - t')$.
$a_{ij}(\boldsymbol{\zeta},t)$ is a function describing the deterministic behavior of $\zeta_{ij}$, and $b_{ij}^{\ell}(\boldsymbol{\zeta},t)$ is a third-rank tensor quantifying how fluctuations in each control parameter affect the friction tensor.  The matching upper and lower indices on $b_{ij}^{\ell}(\boldsymbol{\zeta},t)\xi_{\ell}(t)$ imply a sum over the index $\ell$, accounting for the effects of all control parameter fluctuations on the $i,j$-th component of the friction.  

Following Gardiner~\cite{gardiner}, we make the small-noise perturbative expansion of $\zeta_{ij}(\boldsymbol{\lambda},t)$ in the small parameter $\epsilon$ representing the magnitude of friction fluctuations:
\begin{equation}
\zeta_{ij}(\boldsymbol{\lambda},t) = \zeta_{ij}^{(0)}(\boldsymbol{\lambda},t) + \epsilon\zeta_{ij}^{(1)}(\boldsymbol{\lambda},t) + \epsilon^2\zeta_{ij}^{(2)}(\boldsymbol{\lambda},t) + \cdots \label{appendix_perturbation_expansion} \ .
\end{equation}
Here $\zeta_{ij}^{(0)}(\boldsymbol{\lambda},t)$ is the solution to the deterministic equation $\md_t\zeta_{ij} = a_{ij}(\boldsymbol{\zeta},t)$ (hence independent of fluctuations in $\boldsymbol{\lambda}$), whereas each of the $\zeta_{ij}^{(n)}(\boldsymbol{\lambda},t), n>0$, has stochastic contributions.
$\epsilon\zeta_{ij}^{(1)}(\boldsymbol{\lambda},t)$ is the leading-order correction to the deterministic solution in the limit of weak noise. Expanding $a_{ij}(\boldsymbol{\zeta},t)$ in $\epsilon$, similarly to~\eqref{appendix_perturbation_expansion}, and grouping terms 
with a common power of $\epsilon$, yields the first-order correction to the deterministic approximation in the weak-noise limit, given by the solution to
\begin{equation}
\dot{\zeta}_{ij}^{(1)}(\boldsymbol{\lambda},t) = -u\left( \zeta_{ij}^{(0)}(\boldsymbol{\lambda},t) \right) + b_{ij}^{\ell}\left( \zeta_{ij}^{(0)}(\boldsymbol{\lambda},t) \right)\xi_{\ell}(t)
\label{appendix_first_order_correction} \ .
\end{equation}
Here $u\left( \zeta_{ij}^{(0)}(\boldsymbol{\lambda},t) \right) \equiv -\partial a\left(\zeta_{ij}^{(0)}\right)/\partial\zeta_{ij}^{(0)}$.
We have used the initial condition that $\zeta_{ij}^{(1)}(\boldsymbol{\lambda} = \boldsymbol{\lambda}_0,t=0) = 0$, which is equivalent to all protocols starting at the same point in control-parameter space. \eqref{appendix_first_order_correction} is simply a time-dependent Ornstein-Uhlenbeck process~\cite{gardiner}.

Thus the two factors involving $\zeta_{ij}$ appearing in~\eqref{appendix_expansion_small} are, to first-order in $\epsilon$,
\begin{subequations}
	\begin{align}
	\langle\zeta_{ij}\rangle_{\Omega_t} &= \zeta_{ij}^{(0)}(\boldsymbol{\lambda},t) = \zeta_{ij}(\langle\boldsymbol{\lambda}\rangle_{\Omega_t},t)\label{appendix_zeta_avg} \\
	\delta\zeta_{ij} &= \sum_m\epsilon^m\zeta_{ij}^{(m)} \approx \epsilon\zeta_{ij}^{(1)}(\boldsymbol{\lambda},t) \label{appendix_friction_fuctutation} \ .
	\end{align}
\end{subequations}
\eqref{appendix_zeta_avg} shows that, to order $\epsilon$ in the weak-noise limit, the average friction along the protocol ensemble is the friction along the average protocol. \eqref{appendix_friction_fuctutation} shows that, to lowest order, the fluctuations in the friction can be approximated by the solution to a time-dependent Ornstein-Uhlenbeck process~\eqref{appendix_first_order_correction}. 

We now perform a similar analysis of the control parameter velocity dynamics.
To begin, we assume that---similarly to the generalized friction $\zeta_{ij}$---the dynamics are described by a linear SDE,
\begin{equation}
\dot{\lambda}^i = a_{\lambda}^i(\lambda^i,t)
+ \epsilon b_{\lambda}^i(\lambda^i,t)\xi_i(t)
\label{appendix_control_parameter_sde} \ ,
\end{equation}
where $\xi_i(t)$ is the component of the (vector) white noise process which affects $\lambda^i$.

The average control parameter velocity is
\begin{equation}
\langle\dot{\lambda}^i\rangle_{\Omega_t} = \langle a_{\lambda}^i(\lambda^i,t)\rangle_{\Omega_t} \label{appendix_control_parameter_velocity_avg} \ ,
\end{equation}
where the average is taken over protocol fluctuations at a given time $t$ within the protocol ensemble $\Omega$.   Fluctuations in the control parameter velocity are then
\begin{subequations}
	\begin{align}
	\delta\dot{\lambda}^i &\equiv \dot{\lambda}^i - \langle a_{\lambda}^i(\lambda^i,t)\rangle_{\Omega_t} \\ &= a_{\lambda}^i(\lambda^i,t) - \langle a_{\lambda}^i(\lambda^i,t)\rangle_{\Omega_t} + \epsilon b_{\lambda}^i(\lambda^i,t)\xi_i(t) \label{appendix_control_parameter_velocity_fluctuation} \ .
	\end{align}
\end{subequations}
Again, following Gardiner~\cite{gardiner} we expand $a_{\lambda}^i$ in a small parameter $\epsilon$,
\begin{subequations}
	\begin{align}
	a_{\lambda}^{i}(\lambda^i,t) &\approx a_{\lambda}^{i,(0)} + \epsilon a_{\lambda}^{i,(1)}\nonumber\\
	&= \langle a_{\lambda}^i(\lambda^i,t)\rangle_{\Omega_t} + \epsilon a_{\lambda}^{i,(1)} \label{appendix_fluctuation_control_parameter} \ ,
	\end{align}
\end{subequations}  
from which it follows that $a_{\lambda}^i(\lambda^i,t) - \langle a_{\lambda}^i(\lambda^i,t)\rangle_{\Omega_t} \approx \epsilon a_{\lambda}^{i,(1)}$, where $a_{\lambda}^{i,(1)}$ is the first-order correction to the dynamics of control parameter $\lambda^i$, analogous to $\zeta_{ij}^{(1)}$ in~\eqref{appendix_perturbation_expansion}.  
Fluctuations in the control parameter velocity are thus (to order $\epsilon$) $\delta\dot{\lambda}^i \approx \epsilon \left[ a_{\lambda}^{i,(1)} + b_{\lambda}^{i}(\lambda^i,t)\xi_i(t) \right]$.

With this weak-noise approximation in terms of mean and linear-order fluctuations in $\zeta_{ij}(\boldsymbol{\lambda})$, $\dot{\lambda}^i$, and $\dot{\lambda}^j$, the covariance terms in the excess power expansion~\eqref{appendix_expansion_small} are
\begin{subequations}
	\label{appendix_covariance}
	\begin{align}
	\langle&\dot{\lambda}^i\rangle_{\Omega_t}\langle\delta\zeta_{ij}\delta\dot{\lambda}^j\rangle_{\Omega_t} = \label{appendix_friction_velocity_covariance} \\
	&\epsilon^2\left[\langle\dot{\lambda}^i\rangle_{\Omega_t}\left\langle\zeta_{ij}^{(1)}a_{\lambda}^{j,(1)}\right\rangle_{\Omega_t} +
	\langle\dot{\lambda}^i\rangle_{\Omega_t}\left\langle\zeta_{ij}^{(1)}b_{\lambda}^j\xi_j(t)\right\rangle_{\Omega_t}\right] \ , \nonumber \\
	\langle&\dot{\lambda}^j\rangle_{\Omega_t}\langle\delta\zeta_{ij}\delta\dot{\lambda}^j\rangle_{\Omega_t} = \label{appendix_friction_velocity_covariance2} \\
	&\epsilon^2\left[\langle\dot{\lambda}^j\rangle_{\Omega_t} \left\langle \zeta_{ij}^{(1)} a_{\lambda}^{i,(1)}\right\rangle_{\Omega_t} +
	\langle\dot{\lambda}^j\rangle_{\Omega_t}\left\langle\zeta_{ij}^{(1)}b_{\lambda}^i\xi_i(t)\right\rangle_{\Omega_t}\right] \ , \nonumber \\
	\langle&\zeta_{ij}\rangle_{\Omega_t}\langle\delta\dot{\lambda}^i\delta\dot{\lambda}^j\rangle_{\Omega_t} = \label{appendix_velocity_velocity_covariance} \\
	&\epsilon^2\left[\langle\zeta_{ij}\rangle_{\Omega_t}\left\langle a_{\lambda}^{i,(1)}a_{\lambda}^{j,(1)}\right\rangle_{\Omega_t} +
	\langle\zeta_{ij}\rangle_{\Omega_t}\left\langle b_{\lambda}^ib_{\lambda}^j\xi_i(t)\xi_j(t)\right\rangle_{\Omega_t}\right] \ , \nonumber \\
	\langle&\delta\dot{\lambda}^i\delta\zeta_{ij}\delta\dot{\lambda}^j\rangle_{\Omega_t} = \mathcal{O}(\epsilon^3) \ ,
	\end{align}
\end{subequations}
where for notational simplicity $b_{\lambda}^i(\lambda^i,t) \equiv b_{\lambda}^i$ and the dependence of $\zeta_{ij}^{(1)}$ on $\boldsymbol{\lambda},t$ is suppressed. We henceforth neglect the final $\mathcal{O}(\epsilon^3)$ term.

In the excess power expansion~\eqref{appendix_expansion_small} the friction-control parameter velocity covariance terms~(\ref{appendix_friction_velocity_covariance},\ref{appendix_friction_velocity_covariance2}) are negligible compared to control parameter velocity covariance term~\eqref{appendix_velocity_velocity_covariance} when
\begin{equation}
\frac{\langle \dot{\lambda}^j\rangle_{\Omega_t}}{\langle\zeta_{ij}\rangle_{\Omega_t}} \ll \frac{\left\langle a_{\lambda}^{i,(1)}a_{\lambda}^{j,(1)}\right\rangle_{\Omega_t} + \left\langle b_{\lambda}^i b_{\lambda}^j \xi_i(t)\xi_j(t)\right\rangle_{\Omega_t}}{\left\langle \zeta_{ij}^{(1)} a_{\lambda}^{i,(1)}\right\rangle_{\Omega_t} + \left\langle \zeta_{ij}^{(1)}b_{\lambda}^{i}\xi_i(t)\right\rangle_{\Omega_t}}\label{appendix_first_order_solution_general} \ .
\end{equation}
Substituting \eqref{appendix_zeta_avg} into \eqref{appendix_covariance}, the expansion reduces to 
(9)
from the main text:
\begin{equation}
\langle \mcP_{\rm ex}\rangle_{\Omega_t} \approx \langle\dot{\lambda}^i\rangle_{\Omega_t}\zeta_{ij}(\langle \boldsymbol{\lambda}\rangle_{\Omega_t})\langle\dot{\lambda}^j\rangle_{\Omega_t} + \zeta_{ij}(\langle\boldsymbol{\lambda}\rangle_{\Omega_t})\langle\delta\dot{\lambda}^i\delta\dot{\lambda}^j\rangle_{\Omega_t} \label{appendix_excess_power_final} \ ,
\end{equation}
This quantifies the excess work associated with completing an ensemble $\Omega$ of protocols, in terms of the average protocol $\langle\Lambda\rangle_{\Omega}$ (defined by the path taken by $\langle\boldsymbol{\lambda}\rangle_{\Omega_t}$) and the control parameter velocity covariance $\langle\delta\dot{\lambda}^i\delta\dot{\lambda}^j\rangle_{\Omega_t}$ along that path.

To explore the limits in which~\eqref{appendix_first_order_solution_general} holds, we derive an explicit expression for the friction fluctuations $\delta\zeta_{ij}$, given by the solution to the time-dependent Ornstein-Uhlenbeck process from~\eqref{appendix_first_order_correction}:
\begin{equation}
\zeta_{ij}^{(1)}(\boldsymbol{\lambda},t) = \int_0^t b_{ij}^{\ell}\left( \zeta_{ij}^{(0)}(\boldsymbol{\lambda},t')\right)e^{-\int_{t'}^t u\left( \zeta_{ij}^{(0)}(\boldsymbol{\lambda},s) \right)\md s}\md \xi_k(t')\label{appendix_first_order_solution} \ .
\end{equation}
Here we have imposed the initial condition that all protocols begin at the same point in control parameter space, equivalent to $\zeta_{ij}^{(1)}(\boldsymbol{\lambda},t)$ vanishing at the start of the protocol, so the boundary term at $t = 0$ in~\eqref{appendix_first_order_solution} vanishes~\cite{gardiner}.   

If we now consider the case where, throughout the protocol, the trajectories of $\zeta_{ij}(\boldsymbol{\lambda},t)$ are at steady-state in the reference frame which is comoving with the deterministic solution $\zeta_{ij}^{(0)}(\boldsymbol{\lambda},t)$, then the fluctuations about the deterministic value of $\zeta_{ij}(\boldsymbol{\lambda},t)$ are independent of time.  This is the same constraint placed on the stochastic protocols in the ensemble considered in \S~IV B 
of the main text.  $u\left( \zeta_{ij}^{(0)}(\boldsymbol{\lambda},t)\right)$ represents a time-dependent variation in the first-order correction about the deterministic limit of~\eqref{appendix_general_linear_sde}, so this constraint requires that $u\left( \zeta_{ij}^{(0)}(\boldsymbol{\lambda},t) \right) = 0$.
Hence at steady state the exponential term in the integral expression~\eqref{appendix_first_order_solution} becomes unity. 

If we also assume that the diffusion tensor for the $\zeta_{ij}$ dynamics is a constant $b_{ij}^{\ell}\left(\zeta_{ij}^{(0)}(\boldsymbol{\lambda},t')\right) = b_{ij}^{\ell,(0)}$, then $\zeta_{ij}^{(1)}(\boldsymbol{\lambda},t)$ is independent of $\boldsymbol{\lambda}$ and the integral expression simplifies greatly:
\begin{subequations}
	\begin{align}
	\zeta_{ij}^{(1)}(t) &\approx b_{ij}^{\ell,(0)}\int_0^{t}\md \xi_{\ell}(t) \\ 
	&= b_{ij}^{\ell,(0)} \xi_{\ell}(t) \label{appendix_friction_fluctuation_final}\ ,
	\end{align}
\end{subequations}
where the final equality follows again from the initial condition that all protocols begin from the same point. 
Subject to these assumptions, we write~\eqref{appendix_covariance} in terms of the parameters of the weak-noise expansion,
\begin{subequations}
	\label{appendix_covariance_final}
	\begin{align}
	&\langle\dot{\lambda}^i\rangle_{\Omega_t}\langle\delta\zeta_{ij}\delta\dot{\lambda}^j\rangle_{\Omega_t} = \epsilon^2 \times \\ 
	&\left[\langle\dot{\lambda}^i\rangle_{\Omega_t}\left\langle a_{\lambda}^{j,(1)}b_{ij}^{\ell,(0)}\xi_{\ell}(t)\right\rangle_{\Omega_t} + \langle\dot{\lambda}^i\rangle_{\Omega_t}\left\langle b_{\lambda}^j\xi_j(t) b_{ij}^{\ell,(0)}\xi_{\ell}(t)\right\rangle_{\Omega_t} \right]\label{appendix_friction_velocity_covariance_final} \nonumber \\
	&\langle\dot{\lambda}^j\rangle_{\Omega_t}\langle\delta\zeta_{ij}\delta\dot{\lambda}^j\rangle_{\Omega_t} = \epsilon^2 \times \\ 
	&\left[\langle\dot{\lambda}^j\rangle_{\Omega_t}\left\langle a_{\lambda}^{i,(1)}b_{ij}^{\ell,(0)}\xi_{\ell}(t) \right\rangle_{\Omega_t} + \langle\dot{\lambda}^j\rangle_{\Omega_t}\left\langle b_{\lambda}^{i}\xi_i(t) b_{ij}^{\ell,(0)}\xi_{\ell}(t)\right\rangle_{\Omega_t}\right] \label{appendix_friction_velocity_covariance2_final} \nonumber \\
	&\langle\zeta_{ij}\rangle_{\Omega_t}\langle\delta\dot{\lambda}^i\delta\dot{\lambda}^j\rangle_{\Omega_t} = \epsilon^2 \times \\ 
	&\left[\langle\zeta_{ij}\rangle_{\Omega_t}\left\langle a_{\lambda}^{i,(1)}a_{\lambda}^{j,(1)}\right\rangle_{\Omega_t} + \langle\zeta_{ij}\rangle_{\Omega_t}\left\langle b_{\lambda}^i b_{\lambda}^j \xi_i(t)\xi_j(t)\right\rangle_{\Omega_t}\right] \label{appendix_velocity_velocity_covariance_final} \nonumber \ .
	\end{align}
\end{subequations}

Substituting \eqref{appendix_covariance_final} and the white noise property $\langle\xi_i(t)\xi_j(t)\rangle = \delta_{ij}$ into \eqref{appendix_first_order_solution_general} gives
\begin{equation}
\frac{\langle\dot{\lambda}^j\rangle_{\Omega_t}}{\langle\zeta_{ij}\rangle_{\Omega_t}} \ll \frac{\left\langle a_{\lambda}^{i,(1)}a_{\lambda}^{j,(1)}\right\rangle_{\Omega_t} + \left\langle \left(b_{\lambda}^{i}\right)^2\right\rangle_{\Omega_t}}{\left\langle a_{\lambda}^{i,(1)}b_{ij}^{\ell,(0)}\xi_{\ell}(t)\right\rangle_{\Omega_t} + \left\langle b_{\lambda}^i b_{ij}^{i,(0)}\right\rangle_{\Omega_t}}\label{appendix_near_average_condition} \ .
\end{equation}
If this holds for all $i,j$ at all points in the average protocol, then the friction-velocity covariance terms in~\eqref{appendix_expansion_small} can be neglected. 
The inequality holds trivially in the asymptotic limit $a_{\lambda}^{i,(1)}\rightarrow 0$ of long protocol durations (slow average control parameter velocities), where \eqref{appendix_near_average_condition} becomes
\begin{equation}
\frac{\langle\dot{\lambda}^j\rangle_{\Omega_t}}{\langle\zeta_{ij}\rangle_{\Omega_t}} \ll \frac{\left\langle \left(b_{\lambda}^i\right)^2\right\rangle_{\Omega_t}}{\left\langle b_{\lambda}^i b_{ij}^{i,(1)}\right\rangle_{\Omega_t}} \ ,
\end{equation}
and the LHS becomes arbitrarily small as $\langle\dot{\lambda^i}\rangle_{\Omega_t}\rightarrow 0$.

\section{\label{appendix:lower_bound}Generalization of lower dissipation bound}

Written explicitly for a single control parameter, our lower bound on excess work~(13)
is
\begin{equation}
\langle W_{\rm ex}\rangle_{\Omega} \geq 2\left\langle\zeta(\langle\lambda\rangle_{\Omega_t})\langle\delta\dot{\lambda}^2\rangle_{\Omega_t}\right\rangle_{\SubScr}^{1/2}\int_{0}^{\tau}\sqrt{\zeta(\langle\lambda\rangle_{\Omega_t})\langle\dot{\lambda}\rangle_{\Omega_t}^2}\md t \label{appendix_lower_bound_1} \ .
\end{equation}
Machta's lower bound on the entropy production of a stochastically driven system is~\cite{machta_2015}
\begin{align}
\langle S_{\rm prod}\rangle &\geq 2\int_0^{\tau}\sqrt{\mathcal{I}(\langle\lambda\rangle_{\Omega_t})\langle\dot{\lambda}\rangle_{\Omega_t}^2}\md t \ ,\label{appendix_lower_bound_machta}
\end{align}
where $\mathcal{I}(\langle\boldsymbol{\lambda}\rangle) = \beta^2\langle\delta f^2\rangle_{\langle\boldsymbol{\lambda}\rangle}$ is the Fisher information matrix.
When control parameter manipulation is the only source of entropy production, and the integral relaxation time $\tau^{\rm R}$ is constant along the protocol, we rewrite \eqref{appendix_lower_bound_machta} as an excess work 
\begin{subequations}
	\begin{align}
	\langle W_{\rm ex}\rangle_{\Omega} &= \kT \langle S_{\rm prod}\rangle_{\Omega} \\
	&\ge 2 \kT \int_0^{\tau}\sqrt{\mathcal{I}(\langle\lambda\rangle_{\Omega_t})\langle\dot{\lambda}\rangle_{\Omega_t}^2}\md t \\
	&= 2\sqrt{\frac{\kT}{\tau^{\rm R}}}\int_0^{\tau}\sqrt{\zeta(\langle\lambda\rangle_{\Omega_t})\langle\dot{\lambda}\rangle_{\Omega_t}^2}\md t \label{appendix_lower_bound_machta_2} \ .
	\end{align}
\end{subequations}
The two lower bounds~\eqref{appendix_lower_bound_1} and \eqref{appendix_lower_bound_machta_2} are thus equivalent when
$\tau^{\rm R} = (\beta\left\langle\zeta(\langle\lambda\rangle_{\Omega_t})\langle\delta\dot{\lambda}^2\rangle_{\Omega_t}\right\rangle_{\SubScr})^{-1}$.

To understand when this equality is achieved, consider Machta's total entropy production (of system and control parameter combined) 
\begin{subequations}
	\begin{align}
	\Delta S_{\rm sys^+} &= \frac{1}{2}\mathcal{I}(\langle\lambda\rangle_{\Omega_t})\Delta\lambda^2 \\ 
	&= \frac{1}{2}\mathcal{I}(\langle\lambda\rangle_{\Omega_t})(\tau^{\rm R})^2\left(\frac{\Delta\lambda}{\tau^{\rm R}}\right)^2 \ ,
	\label{appendix_lower_bound_machta_term}
	\end{align}
\end{subequations}
due to a fluctuation in the protocol (Eq.~9 from~\cite{machta_2015}).
Rewriting this as an excess work
\begin{subequations}
	\begin{align}
	W_{\rm ex}^{\rm stoch} = \frac{1}{2}\zeta(\langle\lambda\rangle_{\Omega_t})\left(\frac{\Delta\lambda}{\tau^{\rm R}}\right)^2\tau^{\rm R}
	\end{align}
\end{subequations}
where we used the decomposition of the generalized friction in~(5) 
, $\zeta(\lambda) = \kT \tau^{\rm R}\mathcal{I}(\lambda)$, and denoted by $W_{\rm ex}^{\rm stoch}$ the contribution to the total excess work due to the stochastic fluctuations of the protocol away from the average path.  For protocols satisfying the locally deterministic limit discussed in 
SI section~V
(and used in the derivation of the lower bound in the main text),
\begin{equation}
\frac{\Delta \lambda}{\tau^{\rm R}} \approx \dot{\lambda} \ . 
\end{equation}
I.e., over time scales comparable to the integral relaxation time $\tau^{\rm R}$, the control parameter velocity is constant.  
This makes
the instantaneous contribution to the excess work due to fluctuations
\begin{equation}
W_{\rm ex}^{\rm stoch} \approx \frac{1}{2}\zeta(\langle\lambda\rangle_{\Omega_t})\dot{\lambda}^2\tau^{\rm R} = \frac{1}{2}\zeta(\langle\lambda\rangle_{\Omega_t})\delta\dot{\lambda}^2\tau^{\rm R} \ ,
\end{equation}
where the final equality expresses that this is the excess work due to a fluctuation away from the average protocol, so $\dot{\lambda} \rightarrow \dot{\lambda} - \langle\dot{\lambda}\rangle_{\Omega_t} = \delta\dot{\lambda}$. 
In a reference frame comoving with the average protocol, if the distribution of control parameter velocities is stationary then the instantaneous probability of observing a trajectory with a particular control parameter fluctuation away from the average is
\begin{equation}
P(\delta\dot{\lambda}) \propto e^{-\frac{\beta}{2}\zeta(\langle\lambda\rangle_{\Omega_t})\delta\dot{\lambda}^2\tau^{\rm R}} \ ,
\end{equation}
and thus the squared fluctuation at each point along the protocol, averaged over the instantaneous ensemble of control parameter velocities, is
\begin{equation}
\langle\delta\dot{\lambda}^2\rangle_{\Omega_t} = (\beta\zeta(\langle\lambda\rangle_{\Omega_t})\tau^{\rm R})^{-1} \ . \label{appendix_lower_bound_average_CP_fluctuation}
\end{equation}  
Rearranging and averaging gives
\begin{equation}
\tau^{\rm R} = (\beta \left\langle\zeta(\langle\lambda\rangle_{\Omega_t})\langle\delta\dot{\lambda}^2\rangle_{\Omega_t}\right\rangle_{\SubScr})^{-1} \ ,
\end{equation}
thereby reducing our general lower bound~\eqref{appendix_lower_bound_1} to Machta's bound~\eqref{appendix_lower_bound_machta}~\cite{machta_2015}.

\section{\label{appendix:simulation-details}Simulation details}
We consider a Brownian particle evolving according to an overdamped Langevin equation,
\begin{equation}
\frac{\md x}{\md t} = -\beta D \, \partial_{x}U(x,\lambda) + \sqrt{2D} \, \xi(t) \ ,
\label{langevin_1}
\end{equation}
where $x$ is the particle's position, $-\partial_x U(x,\lambda)$ is the force experienced by the particle due to the potential $U(x,\lambda)$, $D$ is the diffusion coefficient, $\beta \equiv (\kT)^{-1}$ is the inverse temperature of the heat bath, and $\xi(t)$ is a zero-mean white noise process with $\langle\xi(t)\xi(t')\rangle = \delta(t-t')$.  The particle evolves in a potential consisting of a harmonic trap and a periodic landscape,
\begin{equation}
U(x,\lambda) = \tfrac{1}{2}k\left[ x - \lambda(t) \right]^2 - \tfrac{1}{2}E^{\ddagger}\cos(\pi x) \ ,
\label{harmonic_particle}
\end{equation}
with control parameter the time-dependent minimum $\lambda(t)$ of the harmonic trap, and energy barrier $E^{\ddagger}$ separating adjacent energy minima of the periodic landscape.

\begin{figure}[h!]
	\centering\includegraphics[width=\columnwidth]{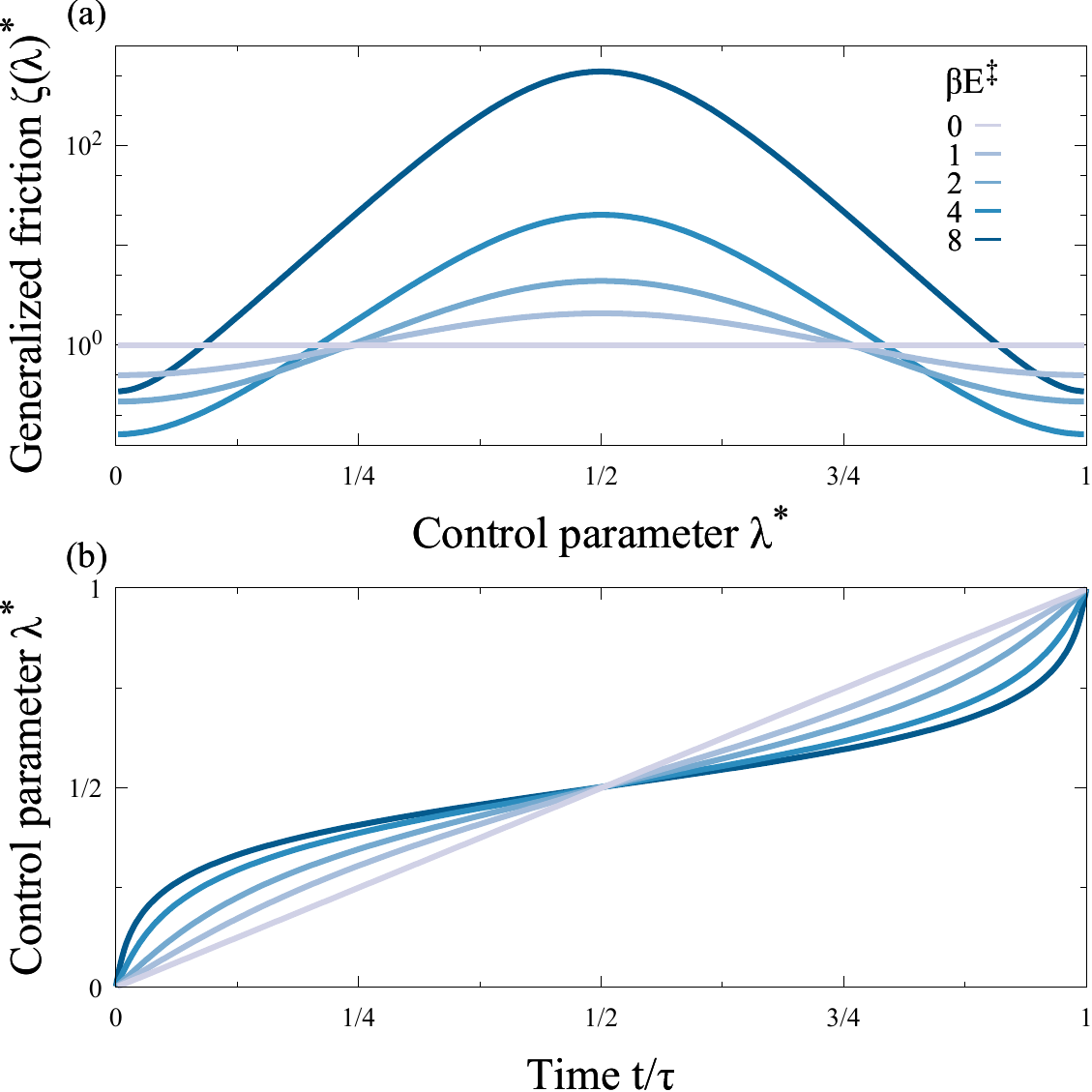}
	\caption{
		\label{fig:appendix_friction}
		{\bf Generalized friction and minimum-work protocols for various periodic barriers.} 
		(a) The generalized friction $\zeta(\lambda)^* = \zeta(\lambda)/\gamma$ is nondimensionalized by the zero-barrier limit $\gamma$, and the friction is periodic over one full period $L$ of the potential. (b) Corresponding minimum-work protocols for $\lambda^* = \lambda/L$ as a function of $t/\tau$.}
\end{figure}

The generalized friction $\zeta(\lambda)$~(16)
is nonuniform over the control parameter landscape when $\beta E^{\ddagger} \neq 0$.
Minimum-work protocols proceed with a control parameter velocity $\dot{\lambda}\propto \zeta(\lambda)^{-1/2}$, with the proportionality fixed by the constraints on the protocol duration (for the periodic-potential ensemble) or the protocol distance (for the stochastic protocol ensemble).
Fig.~\ref{fig:appendix_friction} shows for various barrier heights the generalized friction and corresponding minimum-work protocol.

For each protocol with a particular average velocity $\langle \dot{\lambda}\rangle_{\Lambda}$, the system is initialized in the corresponding periodic steady state for the minimum-work protocol.  
The periodic steady state is achieved when
\begin{equation}
P(x,t+\tau_{\rm L}) = P(x,t) \ ,
\end{equation}
where $\tau_{\rm L}$ is the time take for the harmonic trap to traverse one periodic image of the underlying potential.

\begin{figure}[h!]
	\centering\includegraphics[width=\columnwidth]{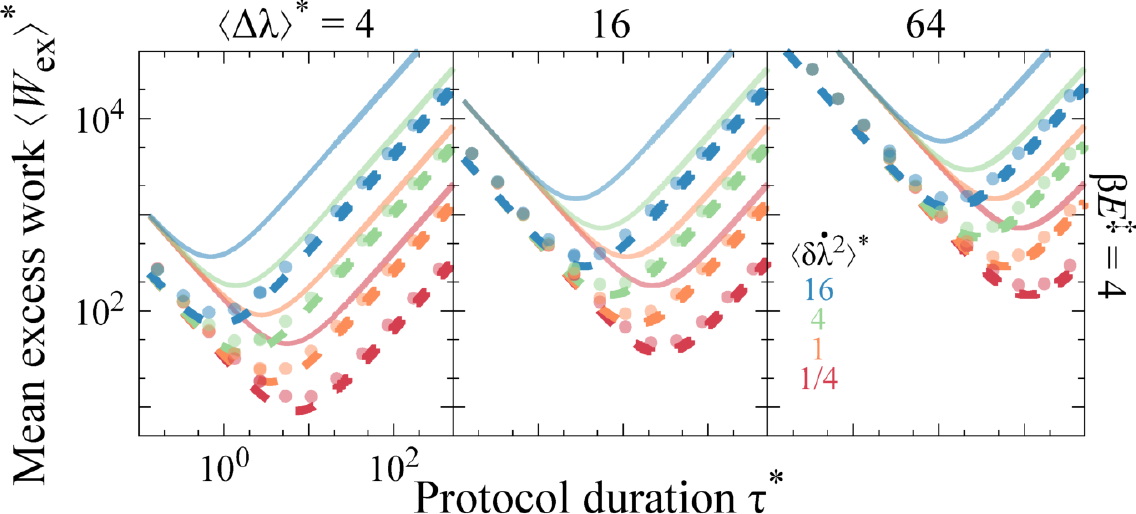}
	\caption{\label{fig:appendix_harmonicwork}
		{\bf Harmonic approximation for excess work agrees with 
			numerical simulations of
			high-barrier periodic-potential ensemble simulations.}  For high-barrier ($\beta E^{\ddagger} = 4$) simulations of the periodic-potential ensemble (dots), the excess work is well approximated by a system driven by a translating harmonic potential (dashed lines), in contrast to the linear-response theory predictions for a system driven by a harmonic trap translating over an underlying periodic potential (solid lines).}
\end{figure}

Numerical results and theoretical predictions in the bottom row ($\beta E^{\ddagger} = 4$) of Fig.~2 of the main text
disagree because in any given protocol ensemble, the average excess work is dominated by the most rapid sampled protocols.
For these fastest protocols, the system position distribution $P(x,t)$ significantly lags the harmonic trap minimum.  As a result, the system experiences a force dominated by the harmonic trap, with little influence from the underlying periodic potential, so the excess work is well-approximated by a system driven solely by a harmonic trap, as confirmed in Fig.~\ref{fig:appendix_harmonicwork}.

In the ensemble of stochastically driven protocols, we consider the zero-barrier potential ($\beta E^{\ddagger} = 0$), where the generalized friction $\zeta(\lambda) = \gamma$ is constant, and thus the minimum-work protocol proceeds with a constant velocity. In this model ensemble, the control parameter $\lambda(t)$ evolves according to the underdamped Langevin equation,
\begin{subequations}
	\label{appendix_SDE}
	\begin{align}
	\frac{\md \lambda}{\md t} &= \dot{\lambda} \label{appendix_SDE_1} \\
	m_{\lambda}\frac{\md\dot{\lambda}}{\md t} &= -\partial_{\lambda}U_{\lambda}(\lambda,\lambda_0) - \frac{1}{\beta_{\lambda} D_\lambda}\dot{\lambda} + \sqrt{\frac{2}{\beta_{\lambda}^2D_{\lambda}}} \, \xi_{\lambda}(t)
	\label{appendix_SDE_2} \ ,
	\end{align}
\end{subequations}
where $D_{\lambda}$ is the diffusion coefficient of the control parameter, $m_{\lambda}$ is the control parameter mass, and $\beta_{\lambda}$ is the inverse temperature of the heat bath in contact with the control parameter.  
The system is initialized in the NESS for the harmonic trap moving with the (constant) average velocity of the protocol ensemble.
The potential energy governing control parameter dynamics is harmonic,
\begin{equation}
U_{\lambda}(\lambda,\lambda_0(t)) = \tfrac{1}{2}k_{\lambda}\left[ \lambda - \lambda_0(t) \right]^2\label{harmonic_CP} \ .
\end{equation}
Averaging the excess work $\langle W_{\rm ex}\rangle_{\Lambda}$ over the protocol ensemble sampled from \eqref{appendix_SDE} gives $\langle W_{\rm ex}\rangle_{\Omega}$~(8).

\section{\label{appendix:const_velocity_ensemble}Exact solution for the zero-barrier periodic potential ensemble}
In the periodic-potential
protocol ensemble, the system dynamics are overdamped and (in the zero-barrier limit [$\beta E^{\ddagger} = 0$]) evolve on a harmonic potential~\eqref{harmonic_particle}.
In this limit, the particle motion obeys an Ornstein-Uhlenbeck process,
\begin{equation}
\frac{\md x}{\md t} = -\beta Dk\left[ x - \lambda(t) \right] + \sqrt{2D} \, \xi(t)
\label{OU_process} \ .
\end{equation}
The control parameter velocity is held constant, $\dot{\lambda}(t) = \dot{\lambda}$, along any protocol realization.  
Direct integration of \eqref{OU_process} for a constant control parameter velocity gives
\begin{align}
x(t) &= x_0e^{-\beta Dkt} \\
&+ \int_0^{t}e^{-\beta Dk(t - s)}\left[ \beta D k\lambda(t) + \sqrt{2D} \, \xi(s) \right]\md s \ . \nonumber
\end{align}
This has mean position
\begin{align}
\langle x\rangle_{\Lambda_t} &= \lambda(0)e^{-\beta Dk t} - \int_0^{t}e^{-\beta Dk(t - s)}\beta Dk\dot{\lambda} \, s \ \md s \ .
\end{align}
Integration by parts gives
\begin{subequations}
	\begin{align}
	\langle x\rangle_{\Lambda_t} &= \lambda(t) - \int_0^{t} e^{-\beta Dk(t - s)}\dot{\lambda} \, \md s \\
	&= \lambda(t) - \frac{\dot{\lambda}}{\beta Dk}\left( 1 - e^{-\beta D k t} \right) \label{expectation} \ .
	\end{align}
\end{subequations}
For long times $t\rightarrow\infty$, the mean asymptotes to
\begin{equation}
\lim_{t\to \infty} \langle x\rangle_{\Lambda_t} = \lambda(t) - \frac{\dot{\lambda}}{\beta Dk}\label{appendix_long_protocol_limit} \ ,
\end{equation}
which lags the trap minimum by a distance $\dot{\lambda}/\beta Dk$~\cite{mazonka_1999}. 

The fluctuating work accumulated for a particular realization of the stochastic process of duration $\tau$ is
\begin{subequations}
	\begin{align}
	W &\equiv \int_0^{\tau} \partial_{t'} U(x,t')\, \md t' \\
	&= -\int_0^{\tau} k\dot{\lambda}[x(t') - \lambda(t')] \md t' \ ,
	\label{fluctuating_work}
	\end{align}
\end{subequations}
for potential $U(x,t')$ taken from \eqref{harmonic_particle}.
The average work accumulated for a protocol of duration $\tau$ is
\begin{equation}
\langle W \rangle_{\Lambda} = -\int_0^{\tau} \, k\dot{\lambda}[\langle x\rangle_{\Lambda_{t'}} - \lambda(t')] \md t' \ .
\end{equation}  
Given that, for the potential considered, there is no free energy change along any protocol, the total work equals the excess work.  

Substituting for $\langle x \rangle_{\Lambda_{t'}}$ using \eqref{expectation} gives the exact mean excess work~\cite{mazonka_1999}
\begin{subequations}
	\begin{align}
	\langle W_{\rm ex} \rangle_{\Lambda} &= k\dot{\lambda}^2\int_0^{\tau}\left[ \int_0^{t'}  \, e^{-\beta D k s} \md s\right]\md t' \\
	&= \frac{\dot{\lambda}^2}{\beta D}\int_0^{\tau}\left[ 1 - e^{-\beta Dkt'} \right]\md t' \\
	&= \frac{\dot{\lambda}^2}{\beta D}\left[ \tau - \frac{1 - e^{-\beta Dk \tau}}{\beta Dk} \right] \ .
	\label{appendix_excess_work_exact_2}
	\end{align}  
\end{subequations}
This differs from the linear-response approximation 
$\langle W_{\rm ex}^{\rm LR} \rangle_{\Lambda} = \dot{\lambda}^2\tau/(\beta D)$ 
((17) for this model)
through the second term representing the exponential decay of the initial transient.

Thus for a system which is initially in the nonequilibrium steady state (NESS) for the protocol velocity $\dot{\lambda}$, the exact mean excess work for a protocol duration of $\tau$ is~\cite{mazonka_1999}
\begin{equation}
\langle W_{\rm ex} \rangle_{\Lambda} = \frac{\dot{\lambda}^2}{\beta D}\tau \label{appendix_excess_work_NESS} \ .
\end{equation}

We now generalize this to randomly choose the (constant) control parameter velocity at the start of each protocol.  
We consider the average of the fluctuating excess work as a path integral over all possible constant-velocity protocols,
\begin{subequations}
	\begin{align}
	\langle W_{\rm ex}\rangle_{\Omega} &= \int P(\lambda_{[0,\tau]}, x_{[0,\tau]})W_{\rm ex}(\lambda_{[0,\tau]}, x_{[0,\tau]}) \nonumber \\
	&\times \mathcal{D}[\lambda_{[0,\tau]}]\mathcal{D}[x_{[0,\tau]}] \\
	&= \int P(\lambda_{[0,\tau]})\mathcal{D}[\lambda_{[0,\tau]}]\\ 
	&\times\int P(x_{[0,\tau]} \mid \lambda_{[0,\tau]})W_{\rm ex}(\lambda_{[0,\tau]},x_{[0,\tau]})\mathcal{D}[x_{[0,\tau]}] \ . \nonumber
	\label{appendix_path_integral}
	\end{align}
\end{subequations}
Here the subscript $[0,\tau]$ indicates that the variable represents a particular realization of the random process on the time interval $[0,\tau]$, and therefore the integral is taken over all paths of the processes $x_{[0,\tau]}$ and $\lambda_{[0,\tau]}$.  The nested integral is the mean excess work for a single protocol $\Lambda$,
\begin{equation}
\langle W_{\rm ex}\rangle_{\Lambda} = \int P(x_{[0,\tau]}|\lambda_{[0,\tau]})W_{\rm ex}(x_{[0,\tau]},\lambda_{[0,\tau]}) \mathcal{D}[x_{[0,\tau]}] \ .
\end{equation}
Thus, the mean work, averaged over all paths $x_{[0,\tau]}$ of the system variable as well as all protocols $\lambda_{[0,\tau]}$, is
\begin{equation}
\langle W_{\rm ex} \rangle_{\Omega} = \int P(\lambda_{[0,\tau]})\langle W_{\rm ex}\rangle_{\Lambda}\mathcal{D}[\lambda_{[0,\tau]}] \ .
\label{appendix_path_integral_final}
\end{equation}

For the ensemble of constant-velocity protocols starting at equilibrium for the initial control parameter value, the excess work $\langle W_{\rm ex} \rangle_{\Lambda}$ is \eqref{appendix_excess_work_exact_2}, and the ensemble average is
\begin{subequations}
	\begin{align}
	\langle W_{\rm ex} \rangle_{\Omega} &= \int P(\lambda_{[0,\tau]})\frac{\dot{\lambda}^2}{\beta D}\left[ \tau - \frac{1 - e^{-\beta D k\tau}}{\beta D k} \right] \mathcal{D}[\lambda_{[0,\tau]}] \\  
	&= \frac{\langle \dot{\lambda}^2\rangle_{\Omega}}{\beta D}\left[\tau - \frac{1 - e^{-\beta D k\tau}}{\beta D k} \right] \label{appendix_exess_work_exact_3} \ .
	\end{align}
\end{subequations}
When the ensemble is initialized in the corresponding NESS, the excess work is \eqref{appendix_excess_work_NESS}, which when averaged over the protocol ensemble gives~(17)
from the main text, 
\begin{equation}
\langle W_{\rm ex} \rangle_{\Omega} = \frac{\langle \dot{\lambda}^2\rangle_{\Omega}}{\beta D} \, \tau \label{appendix_work_exact_NESS_2} \ .
\end{equation}
Both \eqref{appendix_exess_work_exact_3} and \eqref{appendix_work_exact_NESS_2} depend on $P(\lambda_{[0,\tau]})$ only through $\dot{\lambda}$, so in both cases, the same mean excess work is produced for any ensemble with constant velocities chosen from any distribution with a given mean and variance.

\section{\label{appendix:limits_of_linear_response}Limits of linear response}
In order to elucidate more precisely the applicability of the linear-response framework~\cite{sivak_2012}, we derive general conditions under which the theory will hold, as well as discuss the physical context where this limit is achieved.  

In \cite{sivak_2012}, the control parameter velocity within the integrand is approximated by its current value
\begin{equation}
\dot{\lambda}^j(t') \approx \dot{\lambda}^j(t_0) \ .
\label{appendix_velocity_truncation}
\end{equation}
Taking this term outside of the integral in the dynamic linear-response approximation for the excess work~(2), and changing variables $t_0 - t' \rightarrow t^{\prime\prime}$,~(2) becomes the excess power discussed in the main text~(4), and introduced in Ref.~\cite{sivak_2012}: $\langle \mcP_{\rm ex}\rangle_{\Lambda_{t_0}} = \dot{\lambda}^i\zeta_{ij}(\boldsymbol{\lambda})\dot{\lambda}^j$.  

Here we consider the next-order terms for $\dot{\lambda}^j(t')$ and derive conditions under which the Taylor series truncation in \eqref{appendix_velocity_truncation} is valid.  
(To consider the conditions under which the linear-response approximation is generally valid, it would be necessary to consider higher-order response functions as well.) 

Expanding the control parameter velocity $\dot{\lambda}^j$ in~(2) about the time argument gives (to first order in $t_0 - t'$)
\begin{align}
&\langle \mcP_{\rm ex}\rangle_{\Lambda_{t_0}} \approx \dot{\lambda}^i(t_0)\int_{-\infty}^{t_0}\md t' \langle\delta f_i(t_0)\delta f_j(t_0 - t')
\rangle_{\boldsymbol{\lambda}(t_0)} \nonumber \\
&\times \left[\dot{\lambda}^j(t_0) + \partial_{t'}\dot{\lambda}^j\Big|_{t_0}
\left( t' - t_0 \right)\right] \ . 
\label{appendix_linear_response_velocity_expanded}
\end{align}
This expansion is well-approximated by~(4)
when the constraint in~(3) is satisfied, 
generally when the protocol $\Lambda$ is smooth and slowly varying over time scales less then the relaxation time of the conjugate forces. 

The smoothness constraint depends on the protocol velocity, but for deterministic protocols can always be satisfied at low velocities as $\ddot{\lambda}^j$ becomes vanishingly small.
In the context of the present work, we are concerned with protocols $\Lambda$ generated from a stochastic equation of motion.  
For instance, for a control parameter confined to a harmonic trap, the relaxation time is $(\beta D_{\lambda}k_{\lambda})^{-1}$.
If the control parameter is sufficiently underdamped, so that the control parameter velocities remain correlated over times which are long compared to the conjugate force relaxation time of the system, then the protocol is effectively smooth over time scales relevant to (3), and the linear-response approximation is good.

In the context of the harmonic potential, an increase in spring constant $k_{\lambda}$ reduces the conjugate force autocorrelation time and provides a complementary mechanism to reach the appropriate limit.  In the deterministic theory, slowing down the control parameter velocity results in better agreement with theory, whereas when the protocols are generated from a stochastic equation of motion there is an additional requirement that the dynamics are sufficiently underdamped that~(3) holds throughout the protocol.

\section{\label{appendix:equivalence_of_ensembles}Equivalence of ensembles}

According to the theory presented in this letter, the excess work is a function only of the control parameter velocity's mean and variance across the protocol ensemble. The primary constraint imposed on the ensembles is that, for each protocol $\Lambda\in\Omega$ the excess work can be accurately approximated by~(6) 
~\cite{sivak_2012}. Details of the ensemble, such as boundary conditions, do not affect the excess work. 

For instance, the ensemble of stochastic protocols considered in \S~IV~B of the main text
has a Brownian Bridge boundary condition on the protocols: each protocol starts and finishes at the same initial and final control parameter values $\lambda_{\rm i}$ and $\lambda_{\rm f}$, respectively, but has a variable duration~\cite{borodin_2000}.  We alternatively consider the ensemble in which each protocol starts at the same control parameter value $\lambda_{\rm i}$ and has the same duration, but has a variable final position $\lambda_{\rm f}$.  Given that both cases have the same control parameter velocity mean and variance, the theoretically predicted excess work is equal, regardless of the substantially different mathematical procedures necessary to find exact solutions.  
This logic also encompasses the zero-barrier periodic-potential ensemble, where the generalized friction is the same as the stochastic protocols ensembles, the excess work values are predicted to be equal. 
However, it is not possible to draw this equivalence with the nonzero-barrier periodic-potential ensemble, because it has a different average protocol.

\begin{figure}[h!]
	\includegraphics[width=0.5\textwidth]{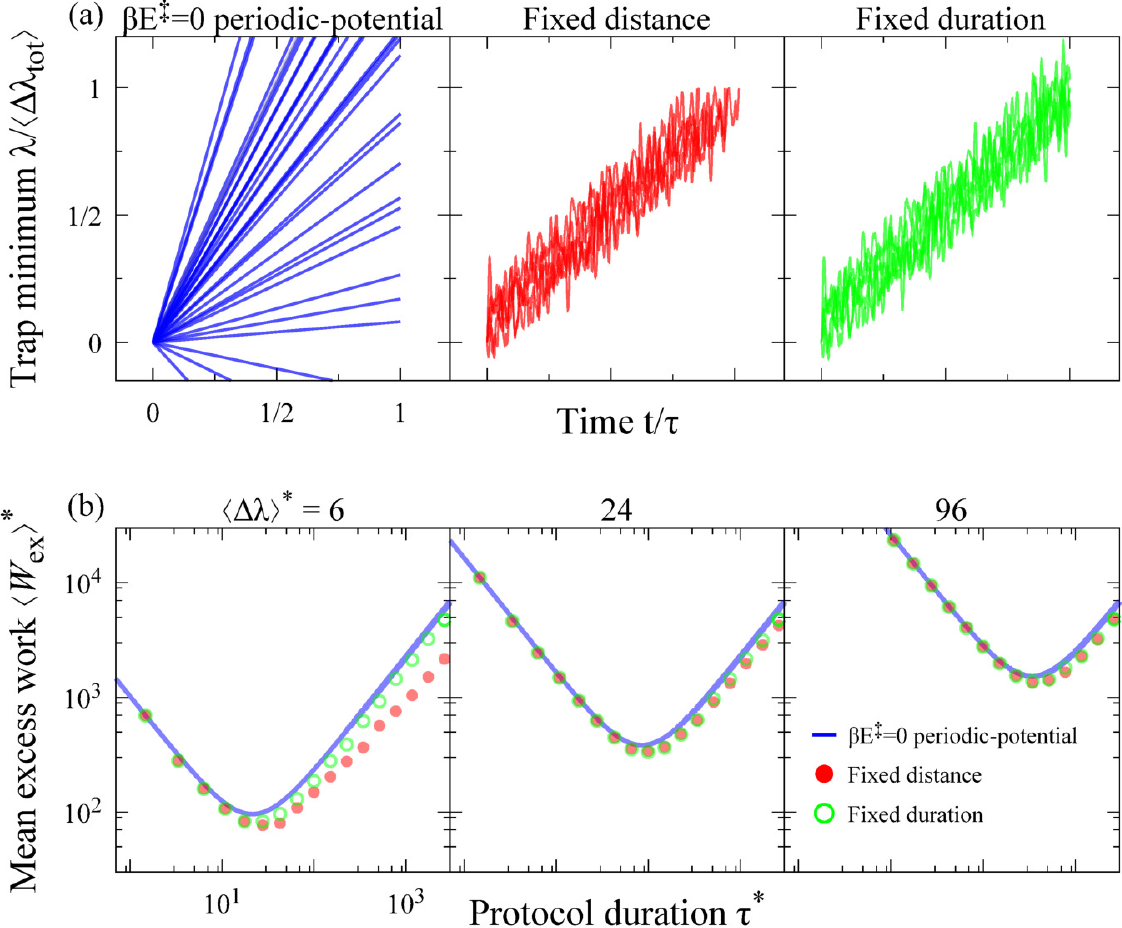}
	\caption{\label{ensemble_compare}\textbf{For near-deterministic protocol ensembles, the excess work depends only on the average protocol and the control parameter velocity variance.}
		(a) Sample trajectories from the zero-barrier ($\beta E^{\ddagger} = 0$) 
		periodic-potential 
		ensemble (left, blue), and from stochastic protocol ensemble with boundary condition of either fixed protocol distance (middle, red) or fixed protocol duration (right, green) with $D_{\lambda}^* \equiv D_{\lambda}/D = 25$. In all cases $\langle\delta\dot{\lambda}^2\rangle_{\Omega}^* = 1$. (b) Average excess work for each ensemble, as a function of protocol duration, with colors matching those in top row. Right: excess work is indeed identical across ensembles in the limits when the linear response~(2) and locally deterministic~\eqref{appendix_protocol_constraint} approximations hold.
	}
\end{figure}

Figure~\ref{ensemble_compare} shows sample trajectories from each of these ensembles with equal theoretically predicted excess works (Fig.~\ref{ensemble_compare}a) and numerically demonstrates the equivalent excess works in the appropriate limit (Fig.~\ref{ensemble_compare}b).  In particular, for large protocol distance (Fig.~\ref{ensemble_compare}b right panel), over all protocol durations the control parameter dynamics satisfy the linear-response~(6)
and locally deterministic~(3) approximations, and hence the different ensembles produce identical mean work that also matches the theoretical approximation derived in the main text~(10).

Interestingly, the statistical moments that appear in~(10)
are those of the control parameter velocity, not its position.  As a result, the precision of the protocol distance, defined as the inverse variance of the final position, can vary significantly depending on the choice of ensemble, while maintaining the same energetic cost.  For instance, the 
periodic-potential protocol ensemble has a precision that decreases secularly with protocol duration, while the stochastically driven protocol ensemble maintains a bound precision regardless of the protocol duration.

\end{document}